# Pedagogical Design Considerations for Mobile Augmented Reality Serious Games (MARSGs): A Literature Review

**Cassidy R. Nelson\* and Joseph L. Gabbard**

Department of Industrial and Systems Engineering, Virginia Tech, Blacksburg, VA ,24060, USA;
cassidynelson@vt.edu, jgabbard@vt.edu

**Abstract**: As technology advances, conceptualizations of effective strategies for teaching and learning shift. Due in part to their facilitation of unique affordances for learning, mobile devices, augmented reality, and games are all becoming more prominent elements in learning environments. In this work, we examine mobile augmented reality serious games (MARSGs) as the intersection of these technology-based experiences and to what effect their combination can yield even greater learning outcomes. We present a PRISMA review of 23 papers (from 610) spanning the entire literature timeline from 2002–2023. Among these works, there is wide variability in the realized application of game elements and pedagogical theories underpinning the game experience. For an educational tool to be effective, it must be designed to facilitate learning while anchored by pedagogical theory. Given that most MARSG developers are not pedagogical experts, this review further provides design considerations regarding which game elements might proffer the best of three major pedagogical theories for modern learning (cognitive constructivism, social constructivism, and behaviorism) based on existing applications. We will also briefly touch on radical constructivism and the instructional elements embedded within MARSGs. Lastly, this work offers a synthesis of current MARSG findings and extended future directions for MARSG development.

## 1. Introduction

Conventional educational approaches are shifting as technology matures and becomes more ubiquitous [1] and new generations of learners become digital natives [1]. One of these shifts has been toward mobile devices as mobile devices afford collaborative, personal, and lifelong learning [2]. Augmented reality (AR), with excellent educational potential [1], contributes to the evolution of information consumption [3], especially in the context of education [4,5]. As learning experiences migrate onto mobile devices and as AR has become more attainable [6], AR has found purchase in the context of mobile augmented reality (MAR) [7,8], with MAR itself bestowing "… situated, authentic, and ubiquitous learning" [3]. Furthermore, due to games' unique ability to engage players in the learning process [9], an additional educational shift has come about through leveraging AR and games for edutainment (mediums that simultaneously entertain and educate) [4].

These shifts in using mobile devices, AR, and learning games are experiencing a correlated, rising prominence and gradual acceptance [3]. This is due both to mobile education's ability to afford game-based learning (GBL), gamification, and serious game (SG) pedagogical theory, as well as AR's ability to deliver valuable pedagogical situations that would be otherwise infeasible [10]. The acceptance of these changes is accelerated by students reporting growing interest in mobile game integration with education [11]. Mobile augmented reality serious games, or, as we term them, MARSGs, are the intersection of these burgeoning areas of pedagogical interest (mobile devices, AR, and learning games) and are uniquely situated to engage inherent student enthusiasm and collaboration [11,12] in the active processes of learning [2] across learning environments and domains [10,13].

Despite these promises, there are currently no pedagogical considerations for effective AR game development that can be leveraged broadly across educational contexts. Thus, MARSG developers often unite pedagogical theory and game elements independently and mostly on an individualized, case-by-case basis. Unfortunately, the resulting literature body features inconsistent realization of pedagogical theory or no pedagogical underpinning whatsoever. This results in little consistency across games and contexts, thus hindering replicability. Further, educational games have better learning outcomes when rooted in learning theory [14] (and to date, many are not).

Wu et al. [15] indicate that design considerations (a contribution of our work presented herein) based on learning theories and empirical evidence for educators and designers would help reconcile tension between the actual reality of educational instruction and the AR game, improving cohesion. Nincarean et al. [16] further posit that pedagogical and learning theories must be a core consideration when developing future applications because an application's educational value is based on more than its inherent features. "To be effective, serious games must incorporate sound cognitive, learning, and pedagogical principles into their design and structure" [17] and must be firmly grounded in educational foundations [14]. Thus, MARSGs should not be built from the bottom-up (i.e., add game elements and determine the learning theories later), but instead be implemented using a top-down approach where designers have already decided what pedagogical theories would best support learning within their domain and, from there, use these considerations to identify potential game elements to support that selected theory.

## Contributions of This Work

To assist MARSG researchers and developers, we present a systematic review of the MARSG literature, distill commonly used educational elements, and offer a set of pedagogically grounded AR design considerations. Concerning the design considerations, it is essential to note that (1) many MARSGs leverage domain-dependent pedagogical theory or learning theories for specific subjects like geometric theory for geometry learning [18]. These domain-dependent theories are captured by our review process but not leveraged in the design considerations. Further, (2) we operationalized game-based learning (GBL) as a pedagogical approach leveraging games [19]. Thus, any MARSG work inherently leverages GBL.

In pursuit of developing design considerations leveraging the aforementioned pedagogical theories, this work examines existing literature on MARSG applications. More specifically, we examined currently leveraged pedagogical theories and the game elements facilitating their unique learning strategies, the instructional elements infused in the game experiences, the game elements taking on a more instructional tone in the context of learning, and how all these components could facilitate cognitive constructivism, social constructivism, and behaviorism.

We expect that our design considerations will further facilitate effective GBL and situated learning by providing a stronger foundation for interweaving domain-dependent theories or situated applications of game elements (thus enhancing games with no theories or only domain-dependent theories). We leveraged three broader, domain-independent pedagogical theories outlined by a review of mobile technologies and learning [2]: cognitive constructivism, social constructivism, and behaviorism. As a brief introduction, (1) behaviorism submits that learning happens through constant feedback in the form of positive and negative stimuli (rewards or punishments) [20], (2) cognitive constructivism posits that knowledge is actively constructed by connecting new knowledge to old [20], and (3) social constructivism hypothesizes that knowledge is constructed through interaction with others [20].

Another contribution of this work focuses on how MARSGs are featured in disparate application domains, making it difficult to discern a synthesized understanding of the trends and conformal relationships of the literature body as a whole. Thus, we additionally offer a census on the breadth of educational domains included in MARSG literature, the diversity of the total participant pools across all included studies, a synthesized understanding of MARSG findings, and extended future directions for the future of MARSGs.

We present these design considerations as a source of inspiration and guidance, encouraging developers to incorporate learning theories as one of their overarching objectives, harmoniously complementing any other game-related goals they may have. This work will allow future developers less familiar with pedagogical theory to 'plug and play' specific game elements more efficiently while giving their projects theoretical grounding. We also advocate for the early integration of pedagogical theories into game design through these guiding principles because we recognize that delving into pedagogical theory can be a challenging endeavor for developers seeking a starting point in this complex domain. Further, this work paints a picture of the trends in the MARSG literature landscape. The following sections outline our working definitions of AR, mobile devices, and serious games. The pedagogical theories are defined more thoroughly in the discussion.

## 2. Related Work

Prior work regarding serious games (SGs) in mobile and AR applications revealed opportunities for further development. Inspiring works of interest were other literature reviews similar to this review, either capturing MARSGs or other partnerships of MARSGs' composite relationships (i.e., MAR, AR games, mobile SGs, etc.).

Yomeldi and Rosmansyah [21] review serious mobile games by focusing on SG features that influence learning outcomes. They find that visualization, interactivity, immersion, and enjoyment significantly influence learning. However, they only included two articles that used AR for game delivery, suggesting the need for a more AR-specific review.

Although AR-focused like ours, Li et al.'s [5] review encompassed all AR serious games, not just those on mobile devices. Like our review, Li et al. examined educational aspects, learning environments, game elements, and outcomes. They found that AR games positively impacted interest, fun, enjoyment, and learning performance. However, their review took a different direction by providing recommendations for AR game design, such as involving learners in the design process, setting clear objectives, promoting social interaction, studying AR features' effects, and understanding game mechanics. While these recommendations were valuable, they did not incorporate pedagogical theory, which is the key aspect of our review.

Sandoval, Koch, and Tizani's [22] review focused on SGs, capturing several device and modality types, including mobile, AR, and virtual reality. Sandoval, Koch, and Tizani identified common characteristics stakeholders desire in games, such as high-quality graphics, rapid content development, interoperability, ubiquity, and immersion. However, they focused on logistical aspects like game engines, devices, and usage rather than the actual game content.

Kosoris, Liu, and Fain's [23] introduced a framework for SGs in 'extended reality,' but it primarily concentrated on virtual reality. Their framework centered on the analysis-driven game and study design, which is essential for future SG development, but from a researcher's perspective, not the player's.

Koutromanos, Sofos, and Avraamidou [13] and Lampropoulos et al. [24] offer the closest works to ours. Koutromanos et al. offer a dedicated review for mobile augmented reality games. However, they do not capture game elements and limit their results to games intended for primary and secondary educational settings. Koutromanos et al. capture educational domains, results, and learning theories. Despite capturing similar characteristics for their data pool, Koutromanos et al. focus on different questions for their work, instead centering on how games are practically embedded within the educational structure and not on individual game elements. Comparatively, Lampropoulos et al. provide an overview of the AR gamification literature space and captured game elements. However, Lampropoulos did not outline underlying learning theory and included 'gamification' works.

In summary, limited work to date focuses on the intersection of serious games in mobile augmented reality [13]. Existing research has concentrated on different aspects of developing these games rather than integrating pedagogical theory into game and instructional elements. Our review aims to provide fresh insights by offering theory-based design considerations, a comprehensive understanding of the literature, and suggesting future directions.

## 3. Background
### 3.1. Augmented Reality

Augmented reality is a real-time (in)direct view of a real environment augmented with virtual information [25], thus supplementing reality instead of replacing it [26]. Worded differently, AR is defined as a "… situation in which a real-world context is dynamically overlaid with coherent location or context-sensitive virtual information" [27]. Hence, AR is the real-time, interactive union of real and virtual content [26]. AR can support pedagogical purposes for gaming platforms because it can furnish "… pedagogically valuable but practically infeasible hypothetical situations" [10] and enhance the information a user can glean from a surrounding environment [1].

Broadly speaking, AR is often delivered via one of two optical means: optical see-through and video pass-through. Due to the inherent nature of current mobile devices as opaque objects, most AR games classically defined as "mobile AR" feature video pass-through AR. AR, as a medium, allows for unprecedented flexibility of physical spaces due to affording instance-based renders, i.e., learners move at their own pace and only see content as they are ready to engage with it, meaning that several learners can move at their own pace within the same physical space. This can be expanded to allow teams of learners to move at different paces as well. AR also allows for unique affordances, like opposing teams occupying the same physical space but only one team being able to see specific content at a time [28]. There are undoubtedly countless unique and creative ways to implement games in AR to facilitate learning that are yet untapped.

### 3.2. Mobile Devices

A core component of mobile learning is that learning moves freely by supporting continuous use across time and location [29]. We define mobile devices as portable, personal, and movable devices that offer individuality, social interactivity, and context-sensitivity [2]. More specifically, this paper operationalizes mobile devices as mobile phones, tablets, laptops, PDAs, and handheld game consoles that afford learning across locations. Thus, we included the entire first quadrant of Naismith et al.'s mobile technologies framework classifications chart [2]. Head-worn displays (HWDs) were not considered 'mobile' for this work. For the date ranges considered in this work, and arguably to some degree still today, HWDs are still too expensive and need longer battery life to be widely used in education when compared to the accessibility of mobile phones and tablets [30].

The MAR component of the MARSGs grants unique game element applications relative to standard computer-based and physical games. Compared to computer-based games, MARSGs offer more organic collaboration aided by the ability of players to physically congregate in interest areas to discuss learning and game content. For example, one game required that players interact with one another to teach each other about viruses and how to defeat them in the game. This game also leveraged a regional leaderboard, creating teamwork for everybody within an area [18], thus creating an incentive to collaborate. Another game allowed players to place world-fixed text hints for future game players [31]. Compared to physical (tangible board and piece) games, competition could be creatively implemented via MARSGs for strategic players trying not to 'give the answer away' to competitors.

### 3.3. Serious Games, Gamification, and Game-Based Learning

The differences between SGs and gamification are nuanced. Both concepts have similar goals to promote learning through game-based thinking and to motivate users [32] and are also core strategies for using games as learning tools [33]. This paper defines *gamification* as adding game elements to an existing educational structure to improve engagement. An example of this is Kahoot! [34], which is a live, competitive quiz featuring points and leaderboards.

This quiz-like game is administered within the traditional classroom structure and behaves like a standard clicker quiz. Clicker quizzes are typically used in situ during lectures and leverage remotes connected to a shared network. Instructors sometimes utilize clicker quizzes for real-time, en masse graded assessments in large lectures. Like clicker quizzes, *Kahoot!* is used during lectures but with game elements [35] like dynamic point awards based on response time and classroom leaderboards. Therefore, *Kahoot!* augments a traditional learning experience with game elements but does not overwrite the traditional learning experience. While powerful, this inherently limits how 'game-ified' the learning experience can be.

Conversely, we define *serious games (SGs)* as an entirely new learning experience provided through a full game [35] with a primary purpose to educate [36] over entertainment. For example, *Minecraft: Education Edition* [37] replaces the standard educational format with a game experience to teach players how to code by providing an in-game environment in which players learn and hone newfound coding knowledge. Moreover, the game can be played at home, thus supporting informal educational environments. SGs offer a more profound departure from traditional classroom strategies due to their inherent nature as dedicated game experiences [35] and functional upending of class exercises. This departure affords less pedagogically constrained game element usage or, worded differently, more freedom to the designer.

Serious games (SGs) are distinct from traditional video games in their primary objective—to educate instead of solely entertain. The core purpose of serious games is to impart knowledge, teach skills, or address specific educational or training objectives, with entertainment serving as a means to engage and motivate learners. There are some classic video games that may feature learning elements, like *Ghost of Tsushima* [38]. *Ghost of Tsushima* includes educational content through ancient artifact collectibles that, once collected, can be viewed as 3D models that have an associated historical fact displayed next to them. However, these artifact collectibles are additional, optional content that is periphery to the core game experience. Consequently, players are entertained first and educated second (and only if they engage with the collectibles hub). SGs feature the reverse with primary aims to "… transfer knowledge pedagogically, in addition to entertainment" [19].

The distinction between SGs and traditional video games is significant given SGs' emphasis on learning outcomes and educational effectiveness. SGs have further been identified as requiring deeper study and understanding [3], thus bolstering our focus on SGs for this review.

Finally, we define *game-based learning (GBL)* as a pedagogical approach leveraging games [19], not a game type itself. Thus, gamification and SGs are the medium through which a developer applies a GBL approach.

### 4. Materials and Methods

We conducted a systematic literature review informed by the Preferred Reporting Items for Systematic Reviews and Meta-Analyses (PRISMA) [39]. The PRISMA framework was adapted for this review to guard against arbitrary decision-making in the conduction of the review, thus attempting to furnish a standard by which to compare the thoroughness of this review to other reviews [40], maintain cohesion, and improve replicability. Adaptation was required as the original PRISMA framework is a medical study framework intended for quantitative reviews, and this paper offers a non-medical, qualitative review. Thus, we employed the following PRISMA elements:

- Title
- Abstract
- Introduction (rationale, objectives)
- Methods (eligibility criteria, information sources, search, study selection, data items)
- Results (study selection, study characteristics, results of individual studies)
- Discussion (summary of evidence, limitations, conclusions)
- Funding (funding disclosure)
  (Included PRISMA items 1–4, 6–9, 11, 17–18, 20, 24–27)

## 4.1. Information Sources

The leveraged databases were Knovel [41], NTIS [42], Compendex [43], Inspec [44], ERIC [45], Web of Science [46], Scopus [47], and IEEE Xplore [48]. Knovel, NTIS, Copnendex, and Inspec were indexed and subsumed under the search interface Engineering Village [49]. The review contained the entire publication span of results (2002 to June 2023) and utilized the keywords augmented reality, serious gam*, and mobile.

## 4.2. Inclusion Criteria

Included articles required the following criteria:

- The papers focused on an application (not technical papers);
- Were in English;
- Were unique (not duplicates);
- Featured a game;
- The game was serious;
- Some participants experienced the game (not a concept paper);
- Participants were evaluated for learning outcomes or their perception of the experience;
- Analysis of participant data was carried out;
- The game was not designed for particular groups (like adults with autism);
- There was an available full text of the paper;
- The paper described employed game elements;
- The game was delivered through augmented reality;
- There was educational content (not purely exercise games)
- The AR was primarily visual;
- The game was furnished through mobile devices.

Worded differently, games were excluded if they were technical papers (like on tracking), were not available in English, were duplicates, did not feature a game, the game was not a serious game (as defined in Section 3.3), there were no participants, the participants were not evaluated for a learning-based outcome or their perception (like only for tracking effectiveness), the game was designed for a specific group (like learners with autism, etc.), there was not an available full text of the paper, the paper failed to describe the game elements utilized, the game was not augmented reality, there was no educational content, the AR was not primarily visual, or the game was furnished through a device that is not mobile (as defined by Section 3.2).

Note that we have excluded games designed for particular groups like learners with autism, ADHD, etc. Given the wide variability in cognition across human development, broad design considerations will already need to be validated within chunked age groups. We pose that special and vulnerable groups should have their own specific, dedicated, and informed reviews that are customized to each group. It is critical not to clump all learners with autism, dyspraxia, dyslexia, ADHD, and so on together because the "average" of all these groups is not likely to provide meaningful findings for any one of these vulnerable groups and in fact might yield harmful generalizations.

## 4.3. Article Sorting

The outlined parameters yielded an initial 610 articles for consideration. Removal of duplicates resulted in 272 remaining records. Application of inclusion criteria to titles and abstracts excluded 138 articles, leaving 134 for closer review. Secondary review yielded a final total of 23 articles for inclusion. Figure 1 outlines the distillation process as well.

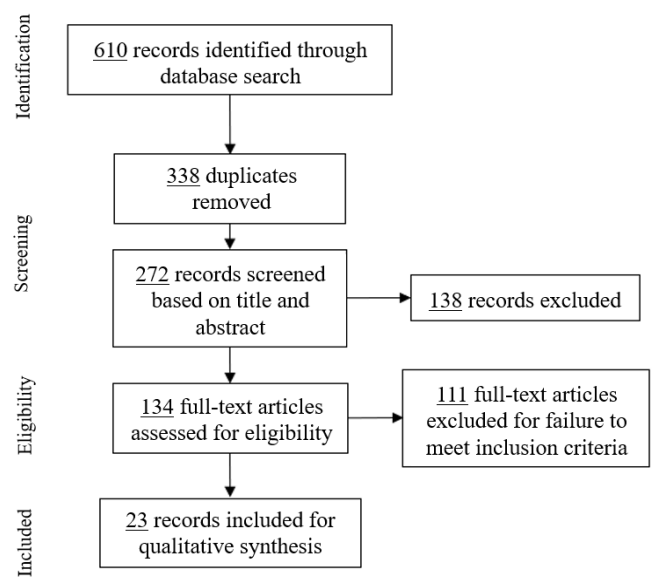

**Figure 1.** PRISMA distillation flow diagram.

## 4.4. Selected Works and Included Data

For reference, the included works are [18,28,31,50–69] and can also be seen listed by citation in Supplementary Table S1. For our purposes, we captured the following data:
1. Supplementary Table S1: Authors, year of publication, number of participants, sample demographics, and country of first author's institution;
2. Supplementary Table S2: Game title, educational subject, game elements, instructional elements, instructional game elements, and the explicitly outlined pedagogical theories leveraged for game design;
3. Supplementary Table S3: Aims/objectives of their study, research design, findings, reported future work, and reported limitations of their work.

## 5. Results
### 5.1. Game Elements

Our review cataloged several game elements discovered in the body of the literature. *Game elements* refer to strategies utilized by game designers to transform the traditional learning experience into a serious game experience. The incident rates of game elements can be seen in Figure 2, while the game element breakdown by citation can be found in Supplementary Table S2.

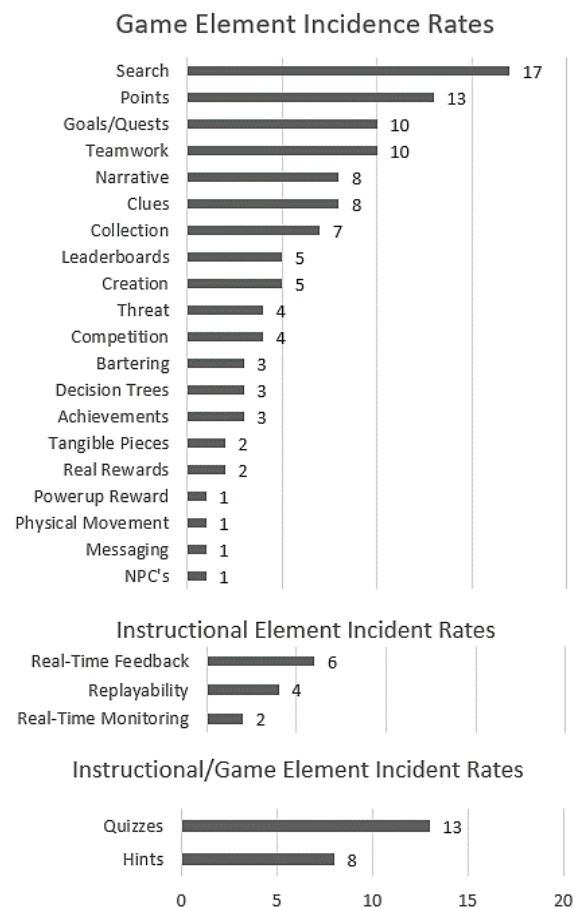

**Figure 2.** Element incident rates.

Every game harnessed the elements of an SG as outlined by Michael and Chen [36]: (1) rules to create and define game parameters for players [70], (2) objectives to assign the player a task, and (3) challenge to intrinsically motivate play [60] and incrementally improve skill or knowledge.

Higher incidence elements were search, points, clues, narrative, teamwork, goals/quests, collection, leaderboards, and creation. Lower incidence game elements included threat, achievements, decision trees, tangible pieces, competition, bartering, messaging, real rewards, physical movement, and non-player characters (NPCs). Each game's incident rate can be found in Figure 2 and defined in this section.

Goals or quests serve as the explicit motivation for the player to play the game [70]. That is, players should engage with the game to help the pirate get treasure [28], etc. However, the goal or quest element's actual aim is to motivate the player to engage with the learning content. Goals denote tasks or challenges, like getting a certain number of points. In contrast, quests are narrative-based, like becoming the best outer space chef via a tournament [61]. *Narratives*, compelling stories driving engagement [60], varied in complexity from simple (like our pirate treasure story) to more involved stories like discovering the secret of a haunted museum [51].

*Search* denotes hunting elements that require players to move around a space to find AR markers (typically QR codes) to unlock the next portion of their game or satisfy their current task. *Clues* were often paired with search in that players must locate these knowledge-based unlocks to unlock previously unknown possibilities. Comparatively, *collection* denotes players needing to 'collect' digital items from their environment to strengthen their character or teach the player something necessary for game progression (like map pieces [51]). These are item-based unlocks. Finally, *creation* as an element is when players must wield their acquired knowledge and/or digital items to create a new item.

*Rewards* were utilized as a means of motivation by leveraging points, achievements, and powerups to motivate players to do their best. *Points* were a quantitatively presented way of measuring performance, while achievements were in-game awards for certain feats. Point mechanics held elevated importance when used to create *competition*, the desire to out-perform oneself or other players through personal bests, leaderboards, or to create threat. For the context of this work, competition only refers to works that explicitly sought to pit players against each other [28], not necessarily where competition could potentially be found in other features like leaderboards. *Leaderboards* display, in rank order, the highest-scoring players with their usernames and scores. This decision was made to denote meaningful distinction about the range of 'competition' in part because for leaderboards to be competitive, players must be intrinsically motivated to perform better than other scorers. The game is not directly asking the players to seek a higher score—instead, it is a passive potential driver. Further, it is possible to use leaderboards to generate camaraderie within teams [57]. *Threat* elicited cautious behavior and was achieved through point penalties for wrong answers (thread avoidance by learning material accurately [63]) or limited time to respond to game queries. *Points* were also employed as currency required to 'buy' hints and powerups (denoted as *bartering*). Points had dynamic presentations, with some games always awarding a consistent amount for correct answers while others awarded scaling point values based on attempts or time taken.

While some games harnessed competition as a core motivator, others implemented collaboration and teamwork instead. Teamwork refers to game elements that foster teamwork between players, like explicitly assigned teams or a shared leaderboard. Some games blended both elements, creating teams that competed against each other [28]. Like competition, collaboration was only denoted as an element for games explicitly seeking to facilitate it. For example, *messaging* was leveraged to create asynchronous player interaction in that players could leave AR messages near interest points in a physical space [31]. Players used this feature to help each other navigate game spaces.

*Decision trees* were subtle and nonobvious to players unless they interacted with other players or replayed the game. Decision trees gave players a dynamic, branching, not-strictly-linear experience, thus creating more freedom in play [54] and enhancing replay value. This means that based on player choice, the narrative or quest objective would 'branch' down a different path, opening some narrative experiences and closing the door on others like specific endings. Thus, decision trees enhanced replayability of the game experience if players wanted to know about the alternative directions of the game. These vary from standard performance checks since a player choosing to help NPC number 1 sends them on a different quest than if they help NPC number 2. *NPC* refers to non-player characters. NPCs are virtual characters that players can interact with to obtain quests or hints and are sometimes the primary vessel through which narratives are delivered.

*Physical movement* refers to using body movement as a game feature beyond simply walking around a space [54]. *Real reward* is a real stakes reward (i.e., extra credit) based on completion and performance [67]. Finally, *tangible pieces* refer to physical boards of game pieces used in conjunction with MAR to progress the game. The relationships between the game-element incident rates and their interplay in the proposed design considerations will be expounded upon in Section 6.3 after we have defined the leveraged pedagogical theories in more detail.

## 5.2. Instructional Elements

Our review also identified instructional elements associated with MARSGs (Figure 2). *Instructional elements* are a core divergence from leisure games. These are features that serve a primarily educational purpose. All game elements were employed in an instructional way. However, instructional elements refer to game features typically not harnessed in classic, standard games. In the context of this review, we found quizzes and real-time feedback to be instructional elements.

*Quizzes* were contextually linked measures of knowledge acquisition and were used to earn/threaten points. However, quizzes are not typically an element of non-educational games. Furthermore, many games offered players constructive feedback on their quiz performance that helped players understand why their answer was incorrect or reinforced correct answers with praise. Thus, *real-time-feedback* was harnessed to ensure players were learning even when performing poorly. Finally, *real-time-monitoring* was a feature meant to facilitate instructor supervision during the learning experience. A breakdown of instructional elements by game can be found in Supplementary Table S2.

## 5.3. Instructional Game Elements

We also classified instructional game elements perched on the line between instructional and game that find unique footing in both worlds (Figure 2). For example, hints and replayability are features typical of leisure games. However, in the context of MARSGs, they have been harnessed for their educational potential.

*Hints* are defined as an instructional and game element because hints aid progress through the game, support learning, and create gentle prompting. This learning-first strategy allowed players to harness their skillsets and knowledge base without simply handing them answers, with well-balanced games consistently situating the players at the edge of their current skills [71]. That is, hints were used primarily for learning, followed by game progression. It is important to note that while not captured by this review, some entertainment-based games use a bartering system with points or monetary currency to 'buy' hints if players get stuck.

Similarly, replayability was considered a game and instructional element as replayability allows players to reinforce their newly acquired knowledge and retry the game to improve their scores (thus fostering competition with themselves). This work only denoted replayability for games that explicitly sought to make their game replayable or motivate the player to replay it. A specific catalog of instructional game elements by game can be found in Supplementary Table S2.

## 5.4. Captured Learning Theories/Frameworks

Lastly, we documented all relevant pedagogical theories and frameworks observed in the included body of literature. All papers utilized GBL, even if not explicitly stated, due to GBL's definition denoting the usage of gaming principles to engage users in learning [32]. Nine papers did not specify any pedagogical grounding for their MARSGs. Two papers specified only domain-dependent theories, with one specifying domain-dependent and independent theories [63]. Examples of leveraged domain-specific theories include geometric theory for mathematical skill development [18], technology threat avoidance theory and protection motivation theory for cybersecurity [63], and biological information theory for viruses [57]. The captured learning theories cataloged by game can be found in Supplementary Table S2.

Nine papers specified only domain-independent theories. These included flow theory [60,64], jigsaw pedagogy [60,64], Gardner's theory of multiple intelligences [65], situated learning, anchored instruction theory [68], inquiry-based learning [64,68], and the constructivist approach (only denoted broadly).

Flow theory and jigsaw pedagogy were utilized twice by two papers introducing and extending the same game [60,64]. Five games sought to facilitate constructivist learning theory [28,51,53,63,64] or spinoff theories rooted in constructivism like inquiry-based learning [67] and authentic learning [56]. Four games leveraged situated learning theory [52,56, 63,64,] or a theory based in situated learning, anchor-based learning [67]. Three other games referenced MARSGs' ability to facilitate situated learning theory but did not indicate how/whether their game explicitly did [57,65,66].

We acknowledge that MARSGs are uniquely empowered to support situated learning (and anchored instruction theory and experiential learning) by anchoring play experiences in relevant physical environments and structuring game tasks around real-world applications of the acquired knowledge [64,72]. Students can learn while grounded in reality [56]. Critically, situated or experiential learning can act as a bridge through which domain-independent and dependent pedagogical theory can simultaneously be applied. More specifically, a game leveraging domain-independent pedagogy can be strengthened by situating itself more firmly in the domain, either through the play experience or by interweaving with domain-dependent pedagogy as domain-dependent pedagogy can be too niche to afford a full game experience in isolation but can enrich an otherwise broad experience.

However, situated learning theory and experiential learning themselves sit on a spectrum of being both domain-dependent and independent. Our design considerations offer a broader, less specific opportunity to infuse situated learning due to the nature of MARSGs being mobile and based in physical reality, often taking place in learning spaces like museums or classrooms. Trending toward more specificity, situated learning theory can also be implemented via any game element with careful, nuanced, and creative design tailored to each domain. However, it is up to each designer and their understanding of the domain to achieve this deeper situated experience. As such, situated learning theory was not explicitly included in the proposed design considerations.

Our design considerations expand on 'the constructivist approach' due to its prominence in the captured literature and its promise for MARSGs as identified by a review of theories for mobile and learning [2]. In addition, our design considerations include behaviorism due to the aforementioned review [2].

## 6. Discussion—MARSG Design Considerations

Informed by the pedagogical theories of cognitive constructivism, social constructivism, and behaviorism, a foundation on which to design MARSGs' game elements emerges. We posit that some game elements are more uniquely situated to enable specific pedagogical strategies over others to contribute to a player's learning (see Figure 3). However, the implementation of each game element should be conducted mindfully, with the pedagogical style and its goals in mind to ensure that players experience the maximum intended benefit. Though we propose that some game elements lend themselves better to specific theories, haphazard application will likely diminish player experiences and learning outcomes. With this in mind, readers should note that we are not positing that game elements explicitly assigned to any pedagogical theory are only usable for that specific theory. In fact, how the games are applied can transform the potential of a game element. This means that it is possible for a game element to have a dual use or purpose, or for learning features of one theory to be present in another. For example, when pursuing points, achievements, or rewards (considered behaviorist elements in Figure 3), the responsibility for learning still rests with the player. Behavioristic learning features can sometimes operate subtly in the background of constructivist learning processes. Finally, we do not intend for developers to feel like they cannot leverage game elements from more than one theory as outlined in Figure 3. Indeed, compelling game/learning experiences may leverage cross-pedagogy game elements. These theories can be integrated to create diverse and effective learning experiences. For instance, educators might use behaviorist principles, such as rewards, to motivate students in cognitive constructivist activities like problem-solving. Social constructivism can enhance both cognitive constructivist and behaviorist approaches through collaborative learning. The relationships between these three theories are outlined in Section 6.2, after they are introduced in Section 6.1.

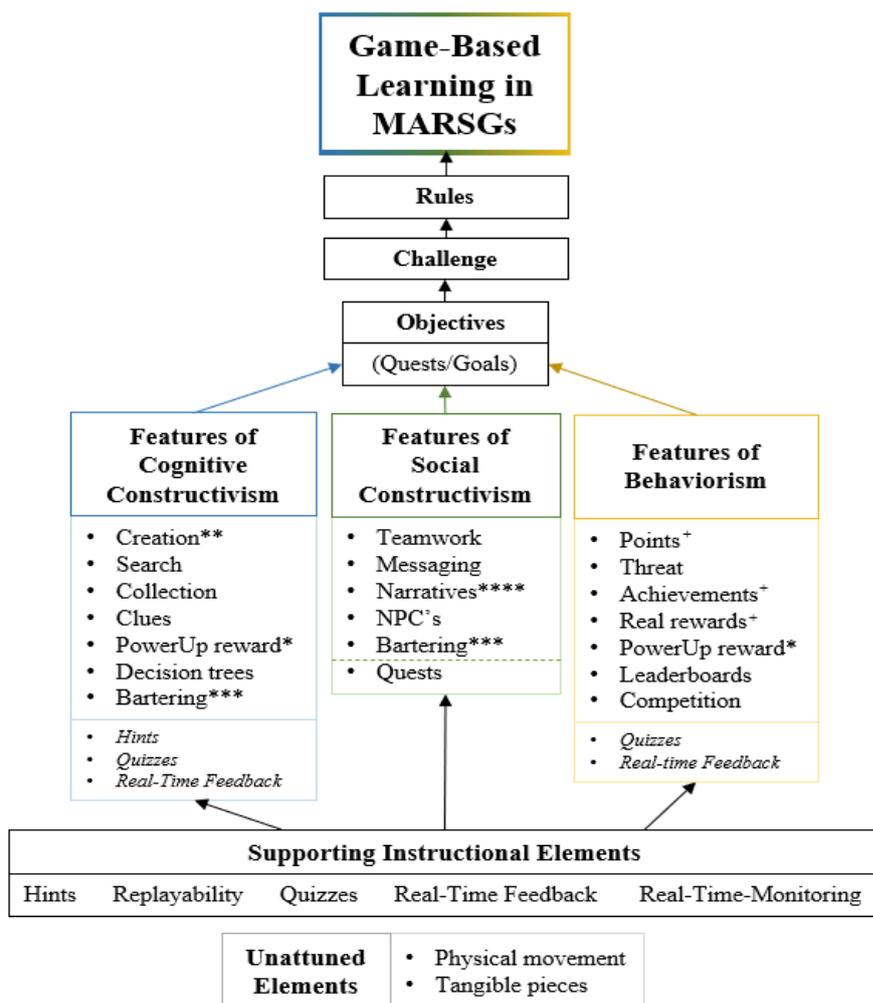

**Figure 3.** MARSG pedagogical design considerations. The *, **, ***, ****, and + symbols denote the flexibility of these game elements to be implemented in other pedagogical strategies if done so thoughtfully. Each of these are talked about specifically in Section 6.1

These considerations offer initial guidance rather than presenting rigid, inflexible rules that must be strictly followed. More specifically, the framework outlined in Figure 3 is a suggested starting point for those unfamiliar with pedagogical theory, and this framework is flexible to tailoring based on the specific educational context or the developer's knowledge base. We outline some of the potential flexibility of several game elements in Section 6.1 for reader consideration. The elements outlined for their flexibility are denoted on Figure 3 with *, **, ***, ****, or + (each of which is defined in the continuing discussion below).

Since GBL is the use of gaming principles to engage users in learning [73], all proposed domain-independent theories are subsumed under the overarching theory of GBL in Figure 3 as each pillar can be leveraged (independently or in tandem) to anchor and afford GBL. Thus, developers should decide early which style of pedagogy seems appropriate for their domain and the complexity of the learning material. From there, objectives and challenges should be considered in the context of intended interactions within the rules of the game experience. Objectives are the explicitly given motivation for players to engage in the experience [36], while appropriate challenge gives the objectives meaning, thus getting students more even engaged [64]. Quests or goals denote the objectives, and objectives should be appropriately challenging. Quests can be layered with a narrative experience to enhance the narrative (and therefore motivation), whereas goals can be without narrative.

A sample utilization of the design considerations could look like this: A designer will identify a learning area and, based on their intended learning outcomes, select a pedagogical theory. The game elements subsumed under that theory branch provide designers a starting point to build their educational content around, being sure to leverage supporting instructional elements. Designers then develop objectives for the player centered around the intended learning goal and ensure they are appropriately challenging. Finally, designers create rules of play that dictate how and the confines within which players engage with those challenges. Now that the flow of the design considerations (Figure 3) has been explained, the pedagogical theories will be defined in more detail.

## 6.1. Pedagogical Theories Contextualized by the Design Considerations

*Behaviorism*, based on Skinner's operant conditioning theories [74] applied by Thorndike (rooted in Pavlov's theories of classical conditioning [2,75]), claims that students learn through constant feedback or reinforcement. Behaviorism has been the dominant theory in pedagogy for over three-quarters of a century [20]; think of rewards for good grades like stickers. Due to the performance-based nature of behaviorism [76], behaviorism pedagogy can be implemented via points, threat, achievements, real rewards, powerup rewards, leaderboards, and competition (Figure 3). Points and achievements act as feedback for player performance (positive reinforcement). This is particularly true for dynamic point systems that change based on response time and attempts. Threat, when used in a point-penalty sense (punishment), is an additional form of feedback with consequences. Real rewards provide a more tangible, outside-the-game reward for performance. Powerup rewards can be given for good performance and take effect immediately, making the game briefly easier (speed powerup, etc.). This game element is denoted with an asterisk (*) because this element can also be used to facilitate cognitive constructivism if applied differently (outlined in the cognitive constructivism section). Leaderboards provide feedback about individual performance and 'social' feedback about performance relative to peers. Behaviorist principles, such as rewards, points, and achievements (denoted with a +), can be also leveraged to inspire student engagement in activities associated with cognitive constructivism, such as problem-solving and critical thinking. Finally, appropriate levels of competition induce flow and increase drive to perform better [77] via this social-comparative feedback. Given AR's ability to spatially integrate virtual content into real environments, behaviorism rewards in AR can be situated more squarely in the player's environment. This means game elements like achievements can carry greater social weight. Further, learning recall is better in the environment in which content is learned [78], providing AR a unique opportunity to facilitate learning and evaluation. Behaviorism is more prescriptive and meaningful for rote, simple learning (like times tables) as it assumes learning is automatic. However, it is not adequate for complex learning [20].

*Constructivism* seeks to facilitate higher-order thinking that is not adequately addressed by behaviorism (the non-visible cognitive processes). It was first theorized by Dewey[79,80] and asserts that learning environments must enable thoughtful reflection and collaborative social negotiation [81]. In the constructivist perspective, information is not simply transferred to the learner; instead, the learner actively constructs knowledge by forming information structures during the learning process. This involves the learner in interpreting information and creating their own understanding based on prior knowledge and experiences. Many included works referenced 'the constructivist approach' broadly, but constructivism is a continuum [20]. Along this continuum lies three broad categories: cognitive constructivism, social constructivism, and radical constructivism [20,82].

*Cognitive constructivism* is founded on Piaget's theories of child development [83] and posits that people actively construct their knowledge, further stating that reality is determined by experiences as a learner by connecting new information to their existing knowledge base [2,84–86] or from active cognizing. Cognitive constructivism is the strictest side of the constructivist continuum because it presupposes an objective external reality that the learner must perfectly reflect. This denotes the seeking of a 'correct' answer [20]. Regarding cognitive constructivism (Figure 3), creation, search, collection, clues, powerup rewards, and decision trees require players to build upon their current knowledge to play the game. For example, creation requires players to have familiarity with learning concepts to make informed choices about possible combinations of materials to generate beneficial and targeted results [57], like by having learners create their own point-earning questions to challenge the other team [69]. Moreover, clues challenge a player's existing knowledge, ideally at the boundary of their understanding [64], thus encouraging players to expand their knowledge for game progression. Decision trees allow flexibility of play. That is, players can make different choices based on their knowledge, particularly if they are replaying the game with a better understanding of the learning content. Powerup rewards that a player can 'hold onto' gives them an opportunity to seek the ideal opportunity to apply that powerup. This requires them to actively understand the game and learning content to make strategic choices. Powerup rewards have an asterisk (*) because they could also support behaviorism. Creation is denoted with two asterisks (**) because this element lends itself the best to games vying for a more radical constructivist approach. Bartering is denoted with three (***) because deeply considering nuanced resources in the context of the game experience could afford cognitive constructivism, while bartering with players could facilitate social constructivism. AR facilitates cognitive constructivism further by increasing engagement and motivation with learning material [53]. Cognitive constructivism is best for mental structures intended to mimic a knowable reality [20]. Finally, cognitive constructivism can include a social element, but social constructivism focuses on social interaction.

Comparatively, *social constructivism* is grounded by Vygotsky's theories in sociocultural psychology [87] and hypothesizes that knowledge is constructed and connected through interaction with others [2,86]. That is, knowledge results from social and shared experiences. Social constructivism lies between cognitive and radical constructivism on the continuum. Further, Jonassen states that true social constructivism involves social negotiation, not competition [81]. If applying social constructivism, teamwork, messaging, narratives, and NPCs are particularly strong elements for consideration (Figure 3). Teamwork or collaboration puts players in direct contact with one another in real-time. Messaging yields asynchronous interaction with other players. Thoughtfully crafted NPCs could afford a degree of social immersion, particularly if they avoid the uncanny valley effect [88]. Narratives and quests can aid immersion and support the 'realness' of NPCs, further compounding a potential social effect (though narrative can be utilized anywhere, hence the quadruple asterisk (****)). Bartering with other players could also facilitate social constructivism. Though competition seems intuitively social, Jonassen states true social constructivism involves social negotiation, not competition [81]. AR being situated in the real environment (or social environment) fosters opportunities for contextualized conversation [61] and collaboration. Social constructivism works well for instilling an ability to generate viable solutions in the context of social coherence [20] (and not based on one 'correct' answer, placing it closer to radical constructivism). This means elements under social constructivism could be used to bolster cognitive constructivism or radical constructivism depending on how they are applied (explained further in the next paragraph).

   ***Radical constructivism*** denotes the opposite, least strict end of the constructivism continuum. While knowledge is still actively constructed, radical constructivism denotes that an external reality is truly unknowable because all knowledge is mediated by senses not adept at rendering perfect reflections [20] (reality subjective). Therefore a student generates a working model understanding of the material and instead seeks nuanced solutions (not an objectively 'correct' answer as there is no 'objective' truth) [20]. We mention radical constructivism due to its rising prominence in pedagogical spaces and ability to afford even more complex learning. However, in the context of these games, most employed quizzes as knowledge checks and therefore sought objective 'correct' answers. Two works [28,59] featured a higher-order type of creation or evaluation that facilitated a more radical constructivist approach. One work tried to teach computational thinking skills [28], and the other managerial decision-making [59]. That is, both focused on complex learning topics requiring the formation of internal working models, not rote memorization. These works significantly diverged from the others because they afforded game scenarios with more flexibility in play, i.e., not seeking 'one' specific correct answer. Nuanced real-time feedback also works toward radical constructivism. Radically constructivist games can leverage many of the game elements we have denoted in our design considerations, but these applications must be nuanced. Further, while radically constructivist games are feasible, they will likely require more mindful development to afford this meaningful flexibility of play [59] or supervision from an instructor to ensure students are learning material appropriately [28] (in this case, a post-assessment verbal report with a human being). As such, radical constructivism is not featured in the design consideration figure as successful application requires a more individualized, domain-dependent approach with greater oversight.

   In summary of these pedagogical theories, while behaviorism, cognitive constructivism, and social constructivism emphasize different aspects of learning, they are interconnected through their focus on learning processes. Integrating these theories in educational contexts allows for a comprehensive approach to teaching and learning (as outlined in more detail in Section 6.2).

   We included instructional elements in our design considerations as these elements can support any pedagogical strategy if mindfully applied (see Figure 3). We do posit that some instructional elements can be uniquely well situated for specific theories, hence their separate, italicized inclusion under some specific theories (see Figure 3). For example, we posit that hints are uniquely strong for cognitive constructivism because they use clues to gently prompt stuck players, encouraging them to stretch their understanding/ability. Quizzes and real-time feedback are denoted for both cognitive constructivism and behaviorism. Quizzes provide a mechanism for measuring a student's grasp of the 'correct' answer while also providing a mechanism for point achievement. Real-time feedback behaves similarly to and in tandem with quizzes; this nuanced feedback of performance that evaluates/explains the core misunderstanding supports cognitive constructivism, whereas simpler (right/wrong) feedback works for behaviorism.

   Unattuned elements refer to game elements found by this review that were not included in the proposed design considerations. Tangible pieces and physical movement were left unassociated due to their lack of inherent cognitive, social, or behavioral implications. These elements could be used to facilitate any of the pedagogical theories.

## 6.2. How Behaviorism, Cognitive Constructivism, and Social Constructivism Interrelate

   When examining the relationships between these theories, several key connections emerge. Behaviorism and cognitive constructivism, although differing in their approaches, both acknowledge the significance of individual cognitive processes. While behaviorism emphasizes external behaviors and stimuli, cognitive constructivism delves into the intricate cognitive mechanisms that drive these behaviors; thus, points could be used to motivate repetitive learning. Cognitive constructivism lays the foundation for social constructivism, as both theories value individual cognitive processes and internal mental structures. However, social constructivism extends this framework by emphasizing the essential role of social interaction and collaboration in shaping individual understanding, which is why teamwork and messaging might provide a meaningful bridge between these two theories. The interplay between behaviorism and social constructivism becomes evident when considering external rewards and reinforcement, defining a potential use case for a leaderboard or achievement structure. In social constructivism, social interactions and collaborative activities can serve as a form of social reinforcement, encouraging learners to actively engage with and construct knowledge through group interactions. Both theories acknowledge the influence of social factors, albeit through different lenses.

   In practical educational settings, these theories can be seamlessly integrated (when implemented strategically) to create diverse and effective learning experiences. Educators have the flexibility to draw from behaviorist principles, such as rewards, to motivate students to participate in cognitive constructivist activities like problem-solving and critical thinking. Furthermore, the incorporation of social constructivism can enhance both cognitive constructivist and behaviorist approaches by fostering collaborative learning experiences. In conclusion, while behaviorism, cognitive constructivism, and social constructivism emphasize distinct aspects of the learning process, they are interconnected through their shared focus on learning mechanisms. Integrating these theories in educational contexts enables a comprehensive and adaptable approach to teaching and learning, catering to diverse learner needs and preferences. Thus, these design considerations offer a starting point for any one of these theories, as well as some guidance on potential combinations of these theories.

## 6.3. Realized Literature Utilization of Elements

| Citation | Cognitive Constructivism | Social Constructivism | Behaviorism | Unattuned |
|---|---|---|---|---|
| [64] | 1 | 1 | 0 | 0 |
| [56] | 0 | 0 | 4 | 0 |
| [31] | 3 | 3 | 3 | 1 |
| [28] | 3 | 1 | 3 | 0 |
| [68] | 0 | 2 | 1 | 1 |
| [58] | 2 | 1 | 0 | 0 |
| [66] | 2 | 0 | 2 | 0 |
| [52] | 2 | 0 | 3 | 0 |
| [54] | 1 | 1 | 1 | 0 |
| [65] | 3 | 1 | 0 | 0 |
| [50] | 1 | 2 | 2 | 0 |
| [51] | 3 | 1 | 0 | 0 |
| [57] | 3 | 1 | 2 | 0 |
| [61] | 1 | 2 | 0 | 2 |
| [62] | 2 | 0 | 1 | 0 |
| [60] | 2 | 2 | 0 | 0 |
| [69] | 2 | 1 | 1 | 1 |
| [53] | 1 | 0 | 2 | 0 |
| [55] | 1 | 1 | 2 | 0 |
| [63] | 2 | 1 | 1 | 1 |
| [67] | 2 | 0 | 2 | 0 |
| [18] | 3 | 1 | 0 | 1 |
| [59] | 1 | 0 | 2 | 0 |
| TOTAL: | 41 | 22 | 32 | 7 |

**Table 1.** Incidence of pedagogical game elements relative to correlated pedagogical theory, i.e., citation [20] featured one cognitive constructivist element and one social constructivist element.

We have proposed some ties between three learning theories and specific game elements. Using these game elements as indicators, twenty-one works used elements that fall under cognitive constructivism, sixteen employed social constructivist elements, and sixteen harnessed behaviorism elements (see Table 1). Only six used any unattuned elements. [68].

Now that we have defined each pedagogical theory and discussed their interrelationships on a theory level, we can refer back to Figure 2 and gain new insights into the distribution of incident rates for various game elements. Among these elements, those with higher incident rates include search (17), points (13), goals/quests (10), teamwork (10), narratives (8), clues (8), collection (7), leaderboards (5), and creation (5). Notably, only two of these elements are anchored under behaviorism, while the remaining behaviorist elements are situated in the bottom half of the usage spectrum. This leaves three elements in the top half categorized as cognitive constructivist and three as social constructivist. This distribution helps explain why only one work utilized game elements from our proposed behaviorist list. Notice that all unattuned elements are in the bottom half (denoting usage of ≤4 times). We suggest that there are additional factors influencing the distribution of game elements. For example, game experiences solely rooted in behaviorism tend to be less engaging for players, often leaning toward simpler gamification. Furthermore, it's worth noting that search and collection-based "treasure-hunt" games currently dominate the landscape of AR serious gaming experiences. This prevalence can be attributed to AR's distinctive real-scale play space and first-person viewing that enables it to provide this immersive treasure-hunting experience.

We can further evaluate these game element incident rates outlined in Figure 2 and Table 1 to ascertain the combined realization of game elements (Figure 4). Fourteen works combined elements from two learning theories, such as cognitive and social or social and behaviorism, demonstrating the flexibility of these considerations. Seven works used both cognitive and social constructivist elements. This distribution is unsurprising given the volume of works using 'constructivism' broadly. Further, although social constructivism focuses on social elements, cognitive constructivism also contains social elements [20]. Six works used cognitive and behaviorist elements, while one used social and behaviorism. Eight works used all three major outlined theories. Only one work used only behaviorism elements (though used in a cognitive context as outlined as feasible earlier), zero used only social constructivism elements, and zero used only cognitive constructivist elements. This higher rate of combined theory usage indicates there may be greater strength in merging multiple theories to buttress an educational game experience (as outlined in Section 6.2), particularly if the melded theories are cognitive constructivism and social constructivism. This further highlights the adaptable and flexible nature of these design considerations when leveraged thoughtfully.

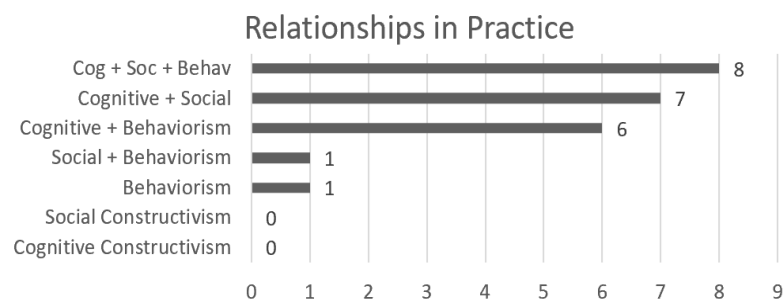

**Figure 4.** The incidence rates of times games used each combination of cognitive constructivism, social constructivism, or behaviorism in tandem, e.g., eight games used all three, and so on.

## 6.4. Contextualizing Impacts of the Proposed Design Considerations

To understand the possible impacts of these design considerations, we offer a synthesis of the included literature body and contextualize our design considerations.

The literature featured a wide array of educational domains (outlined in Supplementary Table S2) facilitated via MARSGs, namely cybersecurity, forensic science, geometry, cultural heritage, local resources, chemistry lab materials, museum content, astronomy, flora/fauna, air quality, viruses, programming, nutritional literacy, and climate change. The two more open-ended games (referenced for radical constructivism) focused on managerial decision-making and computational thinking, that is, complex learning topics without an objective 'right' answer.

Simultaneously, the MARSG literature provided startling participant diversity given that XR research is biased, with 95% of the global population excluded [89] due to an over-reliance on the M-WEIRD [89,90] population (Male, Western, Educated, Industrialized, Rich, and Democratic). Our review found that 1333 players from 13 countries (Australia 7.35%, England 0.38%, Finland 13.2%, Germany 43.98%, Greece 4.2%, Italy 3.3%, Malta 0.45%, New Zealand 1.05%, Peru 1.56%, Portugal 25.13%, Serbia 1.42%, Taiwan (Republic of China) 3.9%, and the United States 34.0%) participated in the included studies. Players were cross-cultural and spanned a wide age range (6–65+). Unfortunately, ~42.84% of the participants' genders were unreported. Of the reported data, ~48.29% of the participants were female and four players (~0.52%) were denoted as 'undefined'. Explicit details organized by game can be found in Supplementary Table S1.

Reported study aims consistently centered on researchers assessing the impact of their own, domain-specific, holistic application of game elements for their MARSGs and not on the learning successes (or failures) of individual game elements. Self-reported perceived phenomena such as flow, engagement, motivation, and effectiveness for domain-specific learning outcomes were measured to capture MARSG impacts. Across objectives, MARSGs had two intended functions: to educate players or to elicit behavioral change through education (one work [55]). Study aims, design, outcomes, limitations, and future directions organized by each included work can be found in Supplementary Table S3.

Overall, players reported feeling highly motivated by and engaged with MARSGs. One study found that after MARSG intervention, players spent more time outside of the MARSG context learning domain content on their own [50]. Motivation and flow were the most measured facets of MARSGs. Players reported feeling that MARSGs added value to the learning experience and that the MARSG modality improved their knowledge [50]. Players also reported higher sociability [31], increased drive for performance [64], and having experienced flow [68]. Additionally, players reported that MARSGs could make learning more enjoyable [65] and that traditional teaching strategies should integrate them [67]. One study found that individual mobile screens did not impede collaboration [28], while another found players reporting immersion and positive affect while playing [57]. Unfortunately, few studies examined the direct learning outcomes resulting from their MARSG interventions. Studies that assessed learning outcomes found that their domain-specific MARSGs significantly impacted players' perceived command of material in pre–post comparisons or that players attained higher subject difficulty by the end of intervention. Encouragingly, one study did achieve pre–post control (traditional learning) vs. intervention group (MARSG) comparisons and found that there was no significant difference in post-learning outcomes between groups [18]. It is hypothesized that this reluctance to evaluate learning outcomes and the non-significance found in the one paper that did is due to the difficulty in ascertaining whether learning outcomes are explicitly from the game experience, whereas other elements like flow and motivation are easier to capture. Finally, all studies examined their games holistically, meaning that no studies examined individual game or instructional elements for their effectiveness. This means that at present, the data is not comprehensive enough to afford predictive claims like xx element (like leaderboards) has a yy effect (5-point increase) on zz (motivation).

All that to say, MARSGs are being implemented in global contexts, across numerous educational domains, for improvement in several facets of learning, from raw outcomes to motivation to learn. The heterogeneity of the player pool demographics and educational domains elucidates the power of MARSGs in both a global and breadth context. The diversity of the study pools provides some theoretical generalizability for our design considerations. Further, the breadth of educational subjects covered by the works included in this paper supports the domain-independent nature of our design considerations. The scale of MARSG implementations speaks to the power MARSGs wield in an educational context to facilitate learning better. Our design considerations provide a vehicle for targeted game element implementation as underpinned by pedagogy. Our design considerations, therefore, stand to have profound implications for future MARSG work, touching numerous cultures, facets of knowledge, and the very study of learning, all by helping future work achieve the requisite pedagogical grounding for maximum effectiveness.

## 6.5. Future Considerations for the MARSG Field Broadly, Contextualized by the Literature

**1. Anchor in theory:** Future studies should always ensure their games are rooted in pedagogical theory. MARSGs will likely be more robust if domain-independent and dependent theories are used in tandem: domain-independent to create a broader learning underpinning and domain-dependent to facilitate the nuance of the learning domain. Future work should also remember that Li et al. recommend close examination of AR features and the study of game mechanics for proper game element selection [5]. Ideally, a pedagogical and domain expert will be involved in MARSG development.

**2. Consider AR's limitations:** Though there are several advantages to using MAR for MARSGs, MAR also experiences unique hindrances to game elements not experienced with physical or computer games. This means MAR is uniquely sensitive to the external learning factors of the environment. Users must have compatible mobile devices and often a degree of network connectivity. MARSGs developers must also be mindful of more than just their game elements and app design. Developers must be conscious that the games will take place, potentially, in varying physical environments. In this scenario, instructors will be the ones to anchor games to a given space. In a poorly adapted physical environment, learners can get lost or frustrated. Moreover, the scale at which participants are required to ambulate around a physical space can limit replayability. Video pass-through refers to using cameras to capture the real world and a screen to display, in real-time, the captured real world with augmented reality graphics superimposed onto the video, thus seemingly re-rendering and overwriting the original world with an updated, enhanced version. When video-pass through AR is delivered through a handheld device, the experience is much like an enhanced "window-on-the-world", (WoW) [91]. Milgram's conceptualization of WoW referred to monitor-based AR (specifically desktop computer monitors) and was considered non-immersive. However, given MAR's inherent mobility and capacity to move around a room with a user and outside of classical classrooms entirely [92], we define MAR as more immersive than the initial conceptualization of WoW AR systems (which were desk-bound). Markerless AR requires players to constantly look through the small 'window' on their mobile device, making identifying and locating markerless content challenging. This is particularly poignant in the context of MARSGs employing the search game element. Thus, boundaries must be defined regarding when players have left the MARSG play area. This additionally means that designers should be mindful of the general limitations of AR when designing their game. Some studies indicated issues specific to their games, like concerns with players not understanding how to use AR [18] and technical issues [61,69].

**3. Engage more players thoughtfully:** Many works indicated concern about their sample sizes as a limitation (see Supplementary Table S3). Future work should ensure that suitable samples are collected with solid research methodology, including control groups and pre–post measures of, at least, learning outcomes. Due to the high degree of relevant factors impacting the effectiveness of learning (i.e., instructor expertise, student expertise, engagement, etc.), large samples are critical to genuinely capturing whether AR provides a significant benefit over and above classical educational strategies. More demographic data should be reported, and gender differences should be considered. Usability studies informed by best practices should also be employed, allowing for practical evaluation and consistency across studies and ensuring learning is not hindered by poorly designed apps. Similar to another review on general MAR games, there are limited studies on MARSGs, resulting in several research areas and questions yet to be explored [13]. Most works only evaluated player perceptions of the game experience (see Supplementary Table S3). Assessment should include more than player self-reports, i.e., objective measures should be captured. Players may feel that they learned more, but whether that is objectively true remains to be seen. Moreover, future work should endeavor to determine the longitudinal effectiveness of learned material. Perhaps there is no significant difference between MARSG and traditional learners at the time of intervention, but there may be significant differences over time given that one study found players spent more time outside the MARSG intervention learning the domain material [50]. As the novelty effect has been shown to increase motivation that subsides with familiarity with the new experience [93], novelty effects should be considered concerning participant enthusiasm surrounding MARSG experiences. Furthermore, longitudinal studies should be conducted regarding maintained learning outcomes and MARSG effectiveness as the technology becomes familiar to players.

**4. Assess implementation barriers:** The success of educational technology integration demands complementary shifts between technological advancement and the educational setting [13,94]. Therefore, barriers associated with MARSG implementation and instructor perceptions of MARSGs should be evaluated. These will impact not only the likelihood of implementation but also the effectiveness of the MARSG if it requires instructor support and setup.

**5. Prediction and validation:** Ideally, a predictive relationship could be abstracted from future work, i.e., predicting the effect of specific game elements on players across measurements of learning, motivation, engagement, fun, difficulty, etc. Future work should validate existing game elements for learning broadly, build upon game element implementations, and create innovative ways to afford learning whilst leveraging the unique affordances of MAR. To do this, future work needs to (1) outline in detail the leveraged game elements and how they were leveraged, (2) evaluate the games for their objective learning effectiveness with pre/post and control group study designs, and (3) strive to understand the unique contributions of each game element to the learning experience. Moreover, more consistent and firm definitions for MARSG components would significantly enhance the development of MARSGs and research regarding each of its component parts.

**6. Transparency:** Unfortunately, many works were excluded from our review because they failed to describe their game in detail (e.g., did not describe specific game elements). Future work should describe the leveraged game elements and how they were utilized such that a reader can understand the gameplay. Moreover, future works should also clearly report demographic data of their participants as ~42.84% of participant genders not being reported is an unacceptable standard.

## 6.6. Limitations of This Work

This review may suffer from selection bias due to limiting included studies to those in English with available full texts accessible through a limited number of electronic databases. Additionally, the selected databases feature a prominent engineering skew. Moreover, game literature is sporadically found across several database types (pedagogical, engineering, art, etc.). There are likely many more games in seemingly obscure places not captured by this review.

As only published texts were reviewed, there could be unpublished data relevant to outlined areas of interest. Compounding this issue is that many MARSGs have been and are being developed by industry groups. Industry groups tend not to publish in academic journals and keep their in-house studies private due to trade secrets; therefore, this review did not capture industry games.

The proposed design considerations are not exhaustive or mutually exclusive. There are multitudes of other pedagogical theories. We selected cognitive constructivism, social constructivism, and behaviorism for the design considerations based on other literature cited in this review and the prominence of constructivism in the selected papers. However, there may be other or additional pedagogical theories not considered for this work to facilitate GBL or that could attune physical movement and real game tokens. Not all learning elements come from the digital learning environment (or game) as the external learning environment can facilitate/hinder the learning experience. MARSGs primarily manifest in formal learning environments, including but not limited to the classroom, e.g., MARSGs extend beyond traditional classrooms and were also prevalent in libraries, museums, guided tours, and parks. MARSGs rarely happened in informal environments like the player's personal neighborhoods or households. However, the external learning environment was not included in this study.

Finally, there is "linguistic chaos" [95] or vague, inconsistent labeling and definitions [95] within each component of MARSGs and, therefore, within the MARSG literature (hence the need for extensive functional definition sections earlier in this work). This chaos impacted our literature search as many papers do not label their games as 'mobile,' instead opting for 'handheld' (which itself has varying meaning) or providing no indication of device type (handheld vs. kiosk, etc.) in the provided keywords. Variance in AR vs. AV on Milgram's spectrum [91], or even nuance within AR across research works [4,6], may also have played a role in the data we attained as there is no definitive consensus for what specific user interface elements or related user experiences constitute a "true" AR experience [4,6]. Furthermore, Sawyer and Smith posit that thirteen terms are used to describe SGs other than the term 'serious games' and include the term gamification. Sawyer and Smith state that these terms additionally vary based on the game's purpose [96]. Worded differently, many games are labeled based on their output, i.e., "social impact game", "persuasive game", etc. [22,96]. These vocabulary difficulties are compounded by the highly interdisciplinary nature of SGs, making it "… impossible for any researcher to be familiar with… every informing discipline" [97]. Articles that did not label their game as mobile, augmented reality, or serious were not captured by this review. This "informing discipline" effect is also key to the sporadic publication venues and databases one can find serious game research, thus aggravating aforementioned issues with database selection.

## 7. Conclusions

This review examines existing MARSG implementations and abstracted themes across game elements and pedagogical theories, then offers a series of design considerations for future MARSG development. Games seeking to leverage cognitive constructivism should consider creation, search, collection, clues, powerup rewards, decision trees, and bartering as potential game elements. Games considering social constructivism should look at teamwork, messaging, narratives, NPCs, and bartering as prospective game elements. Finally, games intending to leverage behaviorism should think about using points, threat, achievements, real rewards, powerup rewards, leaderboards, and competition in their games. Any of these game elements can be used for any of these pedagogical theories if implemented mindfully—these are just some sample clusters to help get designers started.

Further, this work offers some general future considerations for the MARSG literature space broadly. Future work should anchor their games in learning theory, consider AR's limitations during development, engage more (both in objective number and diversity) players thoughtfully, assess implementation barriers for MARSGs broadly, and work toward prediction and validation of game elements.

We found numerous domains supported by MARSGs, with a wide variety of research claims buttressed with greater-than-average participant diversity. Generally, players reported a perceived increase in motivation and enjoyment and that MARSGs were useful. Players likewise reported that they felt the MARSG experience enhanced their learning. Reviewed studies found that MARSG experiences positively impacted player command of learning material, though it is unclear whether this is significantly different from increases seen with traditional learning interventions over the same material. The proposed design considerations could help with consistency across papers to assure replicability and could enhance MARSG effectiveness.

## Future Work for These Design Considerations

A deeper dive into the literature, particularly into more pedagogical and game databases, with the aid of a librarian should be considered. Each game element found by this review and any future creative game element implementations should be validated for effectiveness for, at least, learning outcomes. Further, an evaluation of industry games should be considered. The proposed design considerations should be validated by assessing the effectiveness of chunked groupings (i.e., effectiveness of all behavioral elements, etc.) and considered for HWDs. Furthermore, combining elements from multiple learning theories should be assessed. Most studies employed game elements from multiple theories simultaneously, and this work did not consider the effects of this. Moreover, the included works all employed a bottom-up approach (user-based evaluation) in their assessment phase. Thus a top-down approach (expert-based analytical evaluation) needs to be applied to compare against [98] to ascertain game effectiveness. A survey of academic developers, industry developers, and pedagogical experts should be conducted to evaluate the practicality of implementing, as well as the realized usability of, these design considerations.

Each pedagogical theory we used for the proposed considerations likewise experiences its own unique limitations in its effectiveness. Therefore, each theory's inherent limitations should be carefully considered for future implementations, e.g., behaviorism is criticized for not accounting for free will and internal influences, cognitive constructivism for the difficulty of making abstractions in the initial phases of learning, and social constructivism for its mediation of knowledge through social experiences (lacking objective truth).

This review attempts to provide broad, domain-independent design considerations. However, it is unclear how effective these elements will be across age groups. There are profound differences in cognitive capabilities in children only a few years apart in age. As such, the guideline and its elements need to be validated with high fidelity across age groups, with age groupings guided by developmental science subject matter experts. Moreover, game elements should be explicitly validated across cultures. This work deliberately excluded games meant for specific populations (people with Alzheimer's, autism, etc.). A review dedicated to those populations is a more appropriate approach.